\documentclass[%
 aip,
 amsmath,amssymb,
preprint,%
]{revtex4-1}

\usepackage{graphicx}
\usepackage{dcolumn}
\usepackage{bm}

\usepackage[utf8]{inputenc}
\usepackage[T1]{fontenc}
\usepackage{mathptmx}

\begin{document}

\title{Position-dependent chiral coupling between single quantum dots and cross waveguides}
\author{Shan Xiao}

\author{Shiyao Wu}

\author{Xin Xie}
\author{Jingnan Yang}
\affiliation{Beijing National Laboratory for Condensed Matter Physics, Institute of Physics, Chinese Academy of Sciences, Beijing 100190, China}
\affiliation{CAS Center for Excellence in Topological Quantum Computation and School of Physical Sciences, University of Chinese Academy of Sciences, Beijing 100049, China}

\author{Wenqi Wei}

\author{Shushu Shi}
\author{Feilong Song}
\author{Sibai Sun}
\author{Jianchen Dang}
\author{Longlong Yang}
\affiliation{Beijing National Laboratory for Condensed Matter Physics, Institute of Physics, Chinese Academy of Sciences, Beijing 100190, China}
\affiliation{CAS Center for Excellence in Topological Quantum Computation and School of Physical Sciences, University of Chinese Academy of Sciences, Beijing 100049, China}

\author{Yunuan Wang}
\affiliation{Beijing National Laboratory for Condensed Matter Physics, Institute of Physics, Chinese Academy of Sciences, Beijing 100190, China}
\affiliation{Key Laboratory of Luminescence and Optical Information, Ministry of Education, Beijing Jiaotong University, Beijing 100044, China}

\author{Zhanchun Zuo}
\thanks{Authors to whom correspondence should be addressed: zczuo@iphy.ac.cn and xlxu@iphy.ac.cn}
\affiliation{Beijing National Laboratory for Condensed Matter Physics, Institute of Physics, Chinese Academy of Sciences, Beijing 100190, China}
\affiliation{CAS Center for Excellence in Topological Quantum Computation and School of Physical Sciences, University of Chinese Academy of Sciences, Beijing 100049, China}

\author{Ting Wang}
\author{Jianjun Zhang}

\author{Xiulai Xu}
\thanks{Authors to whom correspondence should be addressed: zczuo@iphy.ac.cn and xlxu@iphy.ac.cn}
\affiliation{Beijing National Laboratory for Condensed Matter Physics, Institute of Physics, Chinese Academy of Sciences, Beijing 100190, China}
\affiliation{CAS Center for Excellence in Topological Quantum Computation and School of Physical Sciences, University of Chinese Academy of Sciences, Beijing 100049, China}
\affiliation{Songshan Lake Materials Laboratory, Dongguan, Guangdong 523808, China}

\date{\today}

\begin{abstract}

  Chiral light-matter interaction between photonic nanostructures with quantum emitters shows great potential to implement spin-photon interfaces for quantum information processing. Position-dependent spin momentum locking of the quantum emitter is important for these chiral coupled nanostructures. Here, we report the position-dependent chiral coupling between quantum dots (QDs) and cross waveguides both numerically and experimentally. Four quantum dots distributed at different positions in the cross section are selected to characterize the chiral properties of the device. Directional emission is achieved in a single waveguide as well as in both two waveguides simultaneously. In addition, the QD position can be determined with the chiral contrasts from four outputs. Therefore, the cross waveguide can function as a one-way unidirectional waveguide and a circularly polarized beam splitter by placing the QD in a rational position, which has potential applications in spin-to-path encoding for complex quantum optical networks at the single-photon level.

\end{abstract}

\maketitle

  Quantum networks based on on-chip photonic integrated circuits have drawn intensive attention for applications in quantum information processing and quantum communication \cite{OBrien2009,xu2016robust,lodahl2017quantum,wang2019}. In such physical platforms, constructing quantum nodes and connecting them with quantum channels have become a fundamental requisite for information storage, processing and exchange \cite{cirac1997quantum,yao2005theory,tanzilli2005photonic,ritter2012elementary,kalb2017entanglement}. So far, various nanophotonic waveguidance schemes have been proposed as channels, such as optical fibers \cite{petersen2014chiral,mitsch2014quantum}, photonic crystals \cite{fattah2013efficient,sollner2015,young2015polarization,mahmoodian2016quantum} and nanobeam waveguides \cite{luxmoore2013,luxmoore2013optical,coles2016,coles2017path,kirvsanske2017indistinguishable,javadi2018spin,mrowinski2019}, which incorporate quantum interfaces transfering information from matter to photonic qubits. In particular, the emergence of chiral interface provides a promising paradigm to deterministically map the quantum state of matter onto the quantum state of the light, thus making the light-matter interaction direction-dependent \cite{lodahl2017}.

 Nanocrystals with intrinsic \cite{ben2011size,ben2013chirality,mukhina2015intrinsic} or induced \cite{cheng2018optically} chirality have been investigated intensively for polarized photon emission. Among of them, self-assembled semiconductor quantum dots (QDs) are excellent candidate for single photon sources with high purity and indistinguishability \cite{xu2004,xu2007,vamivakas2009spin,bracker2006engineering,aharonovich2016}, which have more potential to scale up on chip. The carrier spins in exciton states contained in these two-level systems have been recognized as matter qubits \cite{imamog1999quantum,warburton2013,yoneda2018quantum,norman2018}. Their spin states determine angular momentum on the photon emission resulting from momentum conservation \cite{tang2018,wu2020electron}. In order to enhance and control light-matter interactions at the single-photon level, efforts have been made to integrate QDs into photonic nanostructures \cite{koseki2009monolithic,faraon2010,claudon2010highly,konishi2011circularly,lodahl2015,ding2016demand,androvitsaneas2016charged,chen2018highly,qian2018two,xie2020cavity,Yang2020}. Nanophotonic waveguides are one of key nanostructures which can deterministically route photons. Transversely confinement of light in the waveguide results in a longitudinal field component, exhibiting local circular polarization, thereby allowing the transverse spin angular momentum and the propagation direction being locked \cite{espinosa2016transverse,abujetas2020}. This scheme is a promising approach for realizing chiral light-matter interfaces and constructing quantum spin networks. Unidirectional spin transfer and path-dependent initialization associated with chirality within the waveguide have been intensively studied \cite{lodahl2017}. Most of them mainly focus on the straight transmission of light along the waveguide region to achieve a one-way flow of information. Therefore, the choice of the targeted path for directional photon routing and the optical links between the devices are limited. Recently spin-dependent splitting of light has been demonstrated in a cross waveguide \cite{luxmoore2013,luxmoore2013optical}. However, position-dependent chiral coupling of single QDs has not been investigated, which is highly desired for spin-to-path encoding in developing complex quantum networks.

  Here, we demonstrate position-dependent chiral light-matter interactions using QDs coupled to a cross waveguide. The propagation direction of circularly polarized light is highly sensitive to QD position with numerical simulation. Experimentally, a magnetic field in a Faraday configuration has been applied on QDs, resulting in two non-degenerate states emitting with opposite circular polarizations. Different unidirectional and polarization-deterministic emission into two waveguides are achieved from the QDs embedded in the cross section of the two orthogonal waveguides. The cross waveguide with quantum emitters could be used not only for deterministic delivery of highly directional single photons, circularly polarized beam splitter, but also for deterministic spin-path coding with controlling the quantum dot position.

\begin{figure}
\centering
\includegraphics[scale=0.54]{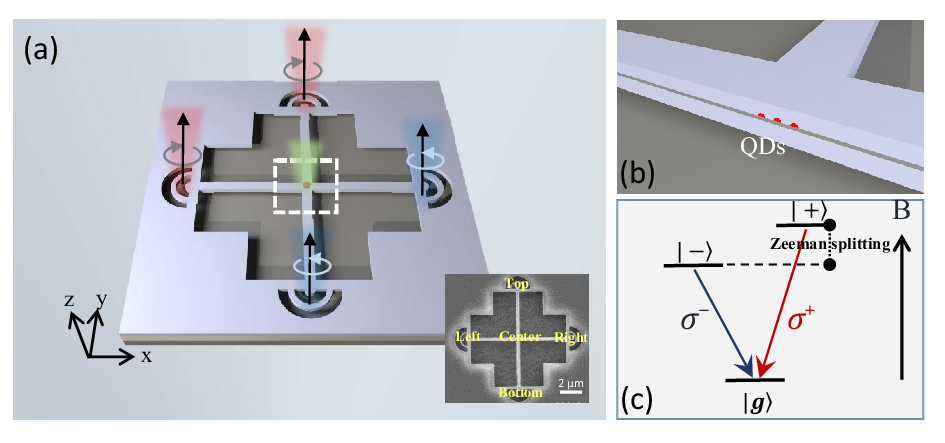}
\caption{(a) Schematic illustration of the cross waveguide and the chiral effect. Inset: SEM image of a fabricated cross waveguide. (b) The xz-cross section of the cross waveguide with QDs embedding in the cross section. (c) QD level structure under a magnetic field in the Faraday configuration, with two circularly polarized exciton transitions  $ {\sigma ^ + } $ and $ {\sigma ^ - } $.  }
\label{F1}
\end{figure}

Figure \ref{F1} (a) illustrates the configuration of the cross waveguide, in which two suspended waveguides are orthogonally positioned and terminated with four circular out-coupling gratings (OCs). The QDs are placed in the cross region [see Fig. \ref{F1}(b)]. A magnetic field in a Faraday configuration is applied on QDs, resulting in a Zeeman splitting of exciton state with two opposite circular polarizations, as shown in Fig. \ref{F1}(c). A wafer composed of a $150 nm$ GaAs membrane with a single layer of InGaAs QDs embedded in the middle has been used to fabricate the devices. The QD density is around $5\times 10^{9} cm^{-2}$. The cross waveguide patterns were transferred to the GaAs layer via electron beam lithography and inductively coupled plasma etching. Thereafter, selective wet etching using hydrofluoric acid was performed to leave the suspended GaAs layer with cross waveguide structure without damaging the optical properties of QDs. A scanning electron microscope (SEM) image of a typical cross waveguide structure with a width of $\sim 280~nm $ is shown in the inset of Fig. \ref{F1}(a).

  The fraction of light coupled into the guided modes is denoted by directional beta factor ($\beta $) , $ {\beta _{\pm}} = {\gamma _{\pm}}/({\gamma _{+}} + {\gamma _{-}} + \Gamma ) $, where $ \gamma _{\pm} $ and $ \Gamma $ are the spontaneous emission rate of photons into one guided mode and all other modes respectively. In the existence of transverse spin component with strongly confined optical fields, the resultant spin momentum locking leads to the asymmetric coupling into the two guided modes, and $ {\beta _{+}} \neq  {\beta _{-}} $. In the extreme case, also called ultra-directional coupling, $ {\beta _{+}} $ approaches 0 and $ {\beta _{-}} $ has a maximum value and vice versa. Additionally, chiral contrast, $ C =  ({\gamma _{-}} - {\gamma _{+}})/({\gamma _{-}} + {\gamma _{+}}) $, is also introduced to quantify the directionality of the emission. By replacing $ {\gamma _{\pm}} $ with $ I_{{\sigma ^ {\pm} }} $ , where $ I_{{\sigma ^ {\pm} }} $ refer to the PL intensity for $ \sigma ^ + $ and $ \sigma ^ - $ polarized emission, the chiral contrast can be measured experimently.

\begin{figure*}
\centering
\includegraphics[scale=0.65]{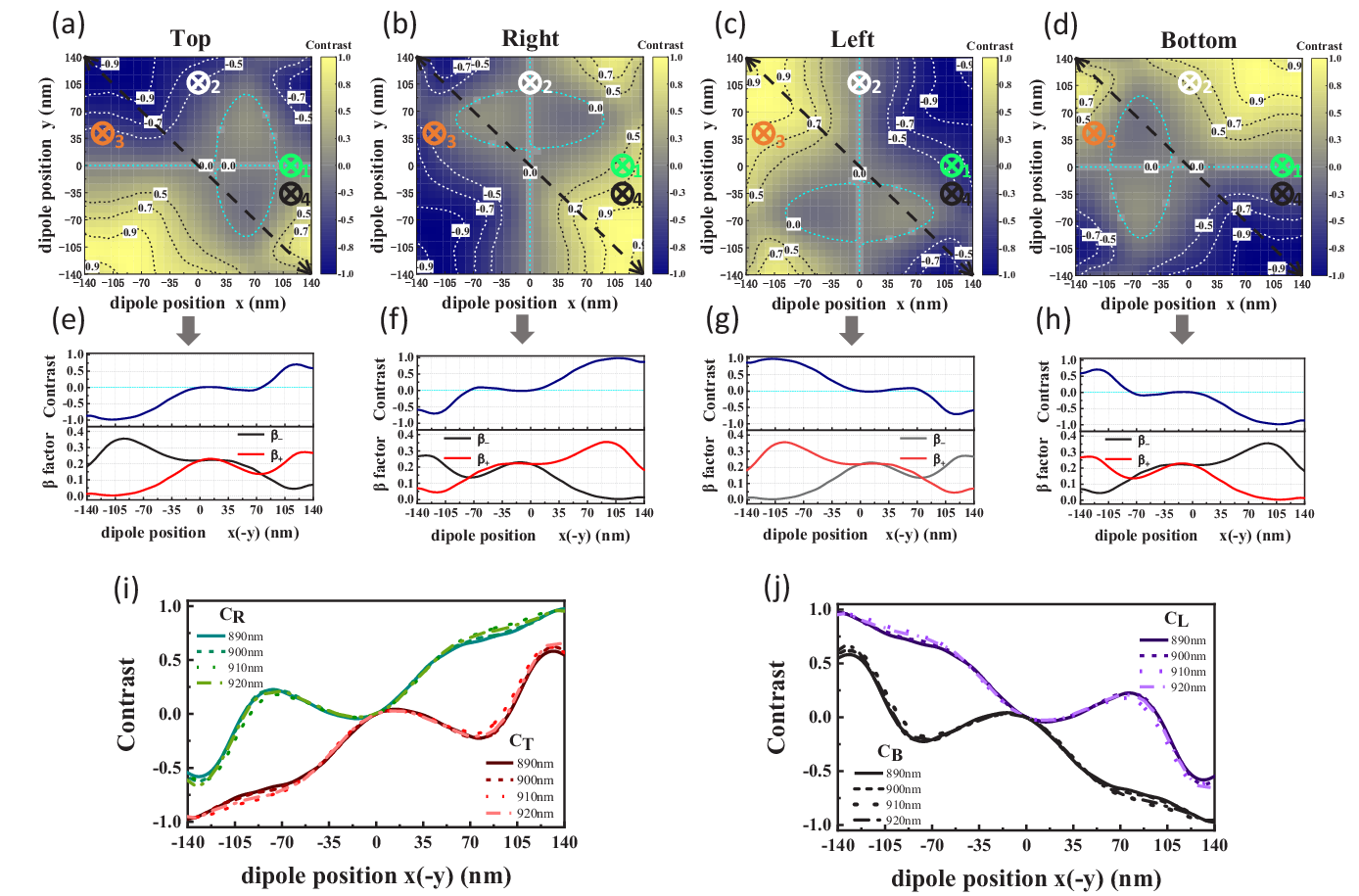}
\caption{(a)-(d) Calculated chiral contrast of four OCs for a right circularly polarized dipole as a function of position in the cross section. The calculation focuses on the fundamental mode of an infinite cross waveguide, with a width of $ 280 nm$, a height of $ 150 nm$ and a wavelength at $920 nm$. The four labels indicate the positions of QDs corresponding to the experimental results. (e)-(h) Calculated chiral contrast and direction $ \beta $ factor as the position of the dipole shifted along the $ -45^\circ $ diagonal direction as shown by the black dashed arrows in (a)-(d). (i) and (j) Calculated chiral contrast for different emission wavelength along the  $ -45^\circ $ diagonal direction.  $\rm C_{R}$ $\rm (C_{L}/C_{T}/C_{B})$ refers to the chiral contrast on Right (Left/Top/Bottom) OC. }
\label{F2}
\end{figure*}

  To investigate the position dependence of the chiral interaction, we calculate $\beta $ and $C$ for a right ($ \sigma ^ - $ ) circularly polarized dipole as a function of its position at the waveguide cross section in the x-y plane with the finite-difference time-domain method. The calculated chiral contrast for all four OCs is shown in Fig. \ref{F2}(a)-(d), all of which can be explicitly mapped with varying chiral contrast distributions. The positive or negative contrast values depend on the position of the dipole and also reflect the polarization of the light output from the four OCs. In reality, the dipole could be positioned at any arbitrary point of the cross section, consequently, highly directional emission of circularly polarized light in the appropriate direction and almost no directional emission can both be achieved experimentally. Since the positions of the dipole with high chiral contrast are concentrated in the four diagonal regions, we therefore extract the values along the $ -45^\circ $ diagonal direction [black dashed arrows in Fig. \ref{F2}(a)-(d)]. Figure \ref{F2}(e)-(h) depict the results of the extracted chiral contrast. For the dipole is positioned close to the center $ x(-y)\approx 0~nm $, the chiral contrast is around 0, i.e. the emitted circularly polarized light is equally transmitted into the four OCs. However, when the dipole is positioned away from the center, such as $ x(-y)\approx 105~nm $, all four OCs obtain a relatively high chiral contrast, implying that the right (left) circularly polarized light is dominantly transmitted into the top/right (bottom/left) OCs. In addition, the coupling efficiency for circularly polarized sources into each OC is also calculated, as shown in Fig. \ref{F2}(e)-(h). One can clearly see the asymmetric emission into the four OCs, thereby inducing directional coupling. Furthermore, we also calculate the chiral contrast for different emission wavelengths along the diagonal direction [see Fig. \ref{F2}(i)and (j)]. The source wavelength is set to between $ 880~nm$ and $ 920~nm$. It can be seen that the contrast is insensitive to the wavelength range of the QD emission with considering the size distribution of QDs. 

To characterize the device spectrally, the devices were cooled down to 4.2 K in a helium bath cryostat equipped with a superconducting magnet and the spatially-selective photoluminescence (PL) measurements were performed with a confocal microscope system with a 0.8 numerical aperture objective lens. A linear polarized continuous-wave laser with a wavelength of $ 532~nm$ was used to non-resonantly excite the QDs in the cross section, resulting in an uncorrelated polarization between absorption and emission. The collected light from the four OCs was then focused on an optical fiber to achieve spatial filtering. The device configuration was imaged by a CCD camera with a background light from a light-emitting diode. The PL spectra were recorded with a single 0.55 m spectrometer. In our case the light spot size is around 1 $\mu m$ and the spectral resolution is about 60 $\mu $eV.  There are about 20 QDs with chiral contrast been observed from fabricated devices. Four QDs (QD1, QD2, QD3 and QD4) selected from the random positions are presented to characterize the performance of the device with a magnetic field of 4 T. All four QDs are embedded in the cross section since they can couple to all four OCs.

\begin{figure}
\centering
\includegraphics[scale=0.24]{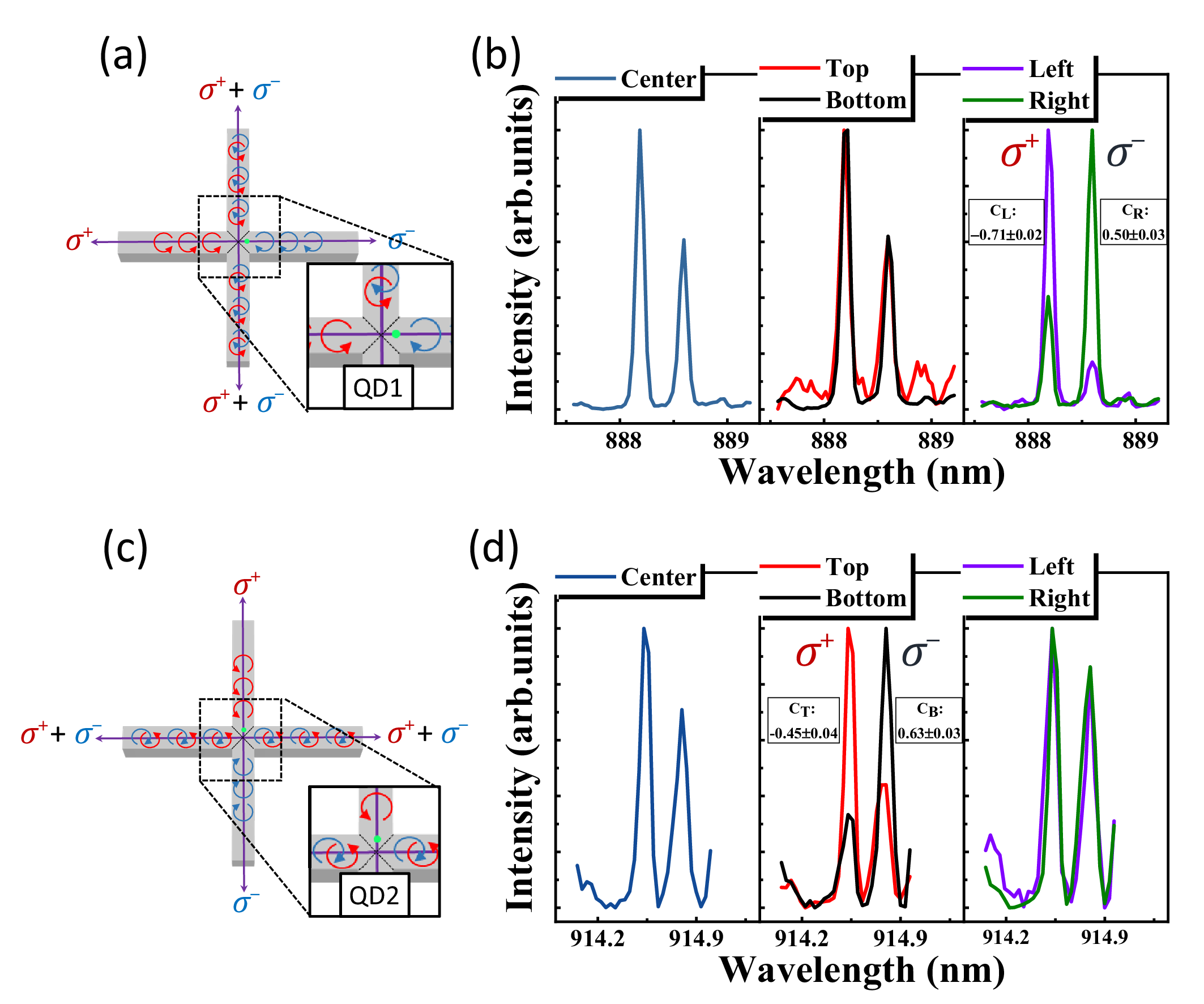}
\caption{(a) and (c) Schematic illustration of the emission direction of QD1 and QD2. The green circles indicate the positions of QDs. (a) The QD1 is chirally coupled to the horizontal waveguide. (c) The QD2 is chirally coupled to the vertical waveguide. (b) and (d) The PL spectra of QD1 (b) and QD2 (d) collected directly from each QD in the center and from four OCs.}
\label{F3}
\end{figure}

Figure \ref{F3}(b) shows the PL spectra of the QD1 collected directly from the QD1 and from the four OCs, respectively. Both two Zeeman-split states corresponding to two opposite circular polarizations are seen vertically from the QD and from the top and bottom outports. However, when we collect the QD spectra from the left and right OCs, significantly asymmetric intensities for the $ {\sigma ^ + } $ and $ {\sigma ^ - } $ polarized transitions are observed, with $ {\sigma ^ + } $ predominantly observed from the left OC and $ {\sigma ^ - } $ predominantly observed from the right OC. This asymmetry is a signature of chiral coupling. We fit the PL emission lines with a Lorentzian function and the intensities can be extracted with errors calculated by error propagation for multi-variables. The chiral contrast is $ -0.71\pm0.02 $ for the left OC and $ 0.50\pm0.03 $ for the right OC. For QD1, spin-dependent directional emission only appears in the horizontal waveguide, while there is no directional emission for the vertical waveguide, as shown schematically in Fig. \ref{F3}(a). In contrast to the transmission results of the QD1, chiral coupling only observed from the vertical waveguide, as shown in Fig. \ref{F3}(c) and (d), with the chiral contrast measured at top OC is $ -0.45\pm0.04 $  and at bottom OC is $ 0.63\pm0.03 $. The divergence of the chiral coupling phenomenon between the two QDs can be attributed to their different positions in the cross section. Referring to the simulations in Fig. \ref{F2}, and the helicity of the circularly polarized light of the OCs, we can infer that the QD1 is located at $ \left( {x,y}  \right) \approx \left( {112,0} \right) nm $ (labelled ``1'') and the QD2 is located at $ \left( {x,y}  \right) \approx \left( {0,112} \right) nm $ (labelled ``2''). Due to the unidirectionality of light from spin-transitions of QDs in one of the waveguides, this device can be functioned as a high efficient one-way waveguide, similar to a single waveguide \cite{coles2016,coles2017path,kirvsanske2017indistinguishable,javadi2018spin,mrowinski2019}.

\begin{figure}
\centering
\includegraphics[scale=0.24]{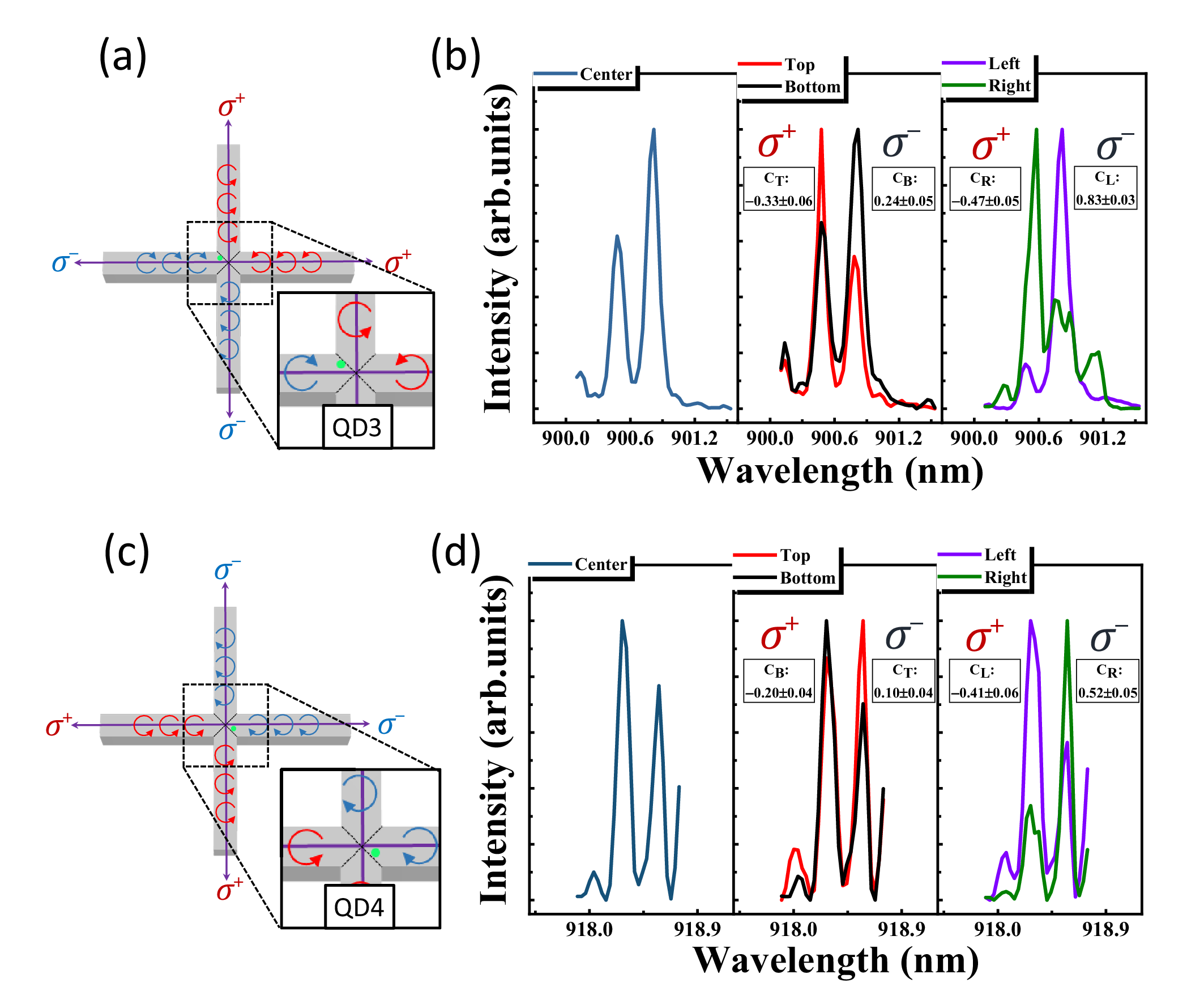}
\caption{(a) and (c) Schematic illustration of the emission direction of QD3 and QD4. The green circles indicate the positions of QDs. (a) The QD3 is chirally coupled to all four OCs, with $ {\sigma ^ + } $ ($ {\sigma ^ - } $) predominantly to the top/right (bottom/left). (c) The QD4 is chirally coupled to all four OCs, with $ {\sigma ^ + } $ ($ {\sigma ^ - } $) predominantly to the bottom/left (top/right). (b) and (d) The PL spectra of QD3 (b) and QD4 (d) collected directly from each QD in the center and from four OCs.}
\label{F4}
\end{figure}

Moreover, chiral coupling can be observed in all four OCs. For QD3, the $ {\sigma ^ + } $ ($ {\sigma ^ - } $) polarized emission is predominantly collected from top/right (bottom/left) OCs [see Fig. \ref{F4}(a) and (b)], with the measured chiral contrast of $ -0.33\pm0.06 $ / $ -0.47\pm0.06 $ ($ 0.24\pm0.05 $ / $ 0.83\pm0.03 $). The direction of the out-coupled circularly light can reverse, as shown in Fig. \ref{F4}(c) and (d) for the PL spectra of the QD4. The $ {\sigma ^ + } $ ($ {\sigma ^ - } $) polarized emission is predominantly collected from bottom/left (top/right) OCs, with the measured chiral contrast of $ -0.20\pm0.04 $ / $ -0.41\pm0.06 $ ($ 0.10\pm0.04 $ / $ 0.52\pm0.05 $). The position of QD3 and QD4 can be inferred at $ \left( {x,y}  \right) \approx \left( {-112,  40} \right) nm $ (labelled ``3'' in Fig. \ref{F2}) and $ \left( {112,-35} \right) nm $ (as labelled ``4'' in Fig. \ref{F2}), respectively. Here, the position-dependence of chiral interaction is confirmed for the two orthogonal waveguides and the function of the device can be extended to a circularly polarized beam splitting. Moreover, the different chiral contrasts from four OCs depending on the position can be used to encode the polarization to different paths with deterministic QD position in the future. With registered QD position precisely \cite{Lee2006,Sapienza2015}, devices with specified spin outputs can be fabricated in a deterministic way. With QDs embedded in a Schottky diode structure or mounted on a piezoelectric thin film actuator, polarization-entangled photon pairs could be integrated in the cross wavegudies with reduced fine structure splitting \cite{Mar2010,Chen2016}. Room temperature chiral coupling devices are also feasible based on the QDs with a high exciton binding energy \cite{ben2013chirality}.

In conclusion, we have demonstrated position-dependent chiral coupling between four QDs and cross waveguides by spatially selective PL measurements. The in-plane transfer can be observed in only one waveguide, with a chiral contrast of up to 0.71, and can also be observed in two orthogonal waveguides, with chiral contrast of up to 0.83. In addition, the change in the in-plane transfer direction of the spin states has been observed. Considering the simulation results, the positions of the four QDs can be determined with measured chiral contrast without considering the reflection from output couplers. Higher chiral contrast could be possible by optimizing the device structure and fabrication process and by precise registration of the QDs positions in the cross section. Spin-to-path encoding could be demonstrated with well defined QD position in the future at the single-photon level. The position-dependent chiral effect can be employed as the basic blocks in integrated quantum optical circuits for quantum networks.

\textbf{Acknowledgements:}

This work was supported by the National Natural Science Foundation of China (Grant Nos. 11934019, 62025507, 11721404, 61775232, 61675228, and 11874419), the Key-Area Research and Development Program of Guangdong Province (Grant No. 2018B030329001), and the Strategic Priority Research Program (Grant No.  XDB28000000) of the Chinese Academy of Sciences.

\textbf{Data availability statement:}

The data that support the findings of this study are available from the corresponding author upon reasonable request.

\nocite{*}

\end{document}